# The Efficacy of Fiction

*or*

# More on the Charge State of Ti in TiO$_2$ and Formal Oxidation States

*Daniel Koch, Sergei Manzhos*[a]

Department of Mechanical Engineering, National University of Singapore

Block EA #07-08, 9 Engineering Drive 1, Singapore 117576

Laser irradiation of titanium can create plasma in which Ti ions with different charge states can be detected, including 0, +1, …+4 |$e$| as well as higher charge states.[1,2] In this way, *absolute* charges can be detected in a straightforward way, as opposed to e.g. XPS on solids where *changes* in charge surrounding the ionic core are detected *indirectly*. Specifically, the Ti$^{4+}$ species in laser ionization experiments is stripped of all of Ti valence electrons and is thereby fully oxidized. Its charge state is +4 |$e$| by definition, but what is its oxidation state? In the context of valence chemistry, it can be no further oxidized and it would be natural to say that it also has an oxidation state of +4. The Ti$^0$…Ti$^{4+}$ species were included in the analysis of ref. 3 where we contrasted them with Ti in TiO$_2$ and argued that the charge state of Ti in TiO$_2$ is close to +3 |$e$| and from there suggested that calling Ti in TiO$_2$ "Ti$^{4+}$" (either as a reference to charge or oxidation state) may be misguided. It is this conclusion that A. Walsh et al. deemed "unphysical" as they rushed to the defense of the concept of formal oxidation states.[4]

The concept of formal oxidation states predates modern quantum and computational chemistry; dating back to 1930s[5], it introduces a set of rules to assign electrons to atoms in a compound based on simple molecular orbital considerations and the ionic approximation.[6] It allows one to walk away from the complexity of continuous electron density and create an easy-to-use scheme with which to characterize matter. Formal oxidation states are fictitious by definition and may more or less reflect the reality. Such a model would be justified if it were (i) necessary to understand phenomena as well as (ii) useful in harnessing them for purposes of humans. In the absence of modern quantum chemistry tools it had likely been a necessary crutch. We are not concerned here whether it continues to be necessary and useful for the rationalization of some phenomena, but specifically consider materials like $TiO_2$ (stoichiometric transition metal oxides). In ref. 7, P. Wolczanski used charge reporter molecules to characterize charge states of metal atoms in multiple transition metal containing compounds; it was found that the charge states are remarkably similar among complexes which correspond to very different formal oxidation states (e.g. about +2 for Fe for compounds where iron is assumed to be from Fe(II) to Fe(IV)). Specifically for Ti, a charge state of about +3 |$e$| was estimated. This follows an earlier work by Raebiger, Lany, and Zunger[8] that arrived at a similar conclusion, namely, that the charge on a transition metal in a compound is very stable with respect to changes of formal oxidation state due to the so-called negative-feedback charge regulation. Already in 2000, Christensen and Carter reported significant covalent character of bonds in $ZrO_2$ and concluded that in $ZrO_2$ "Zr is likely to be Zr(II) like" in spite of the formal oxidation state of Zr(IV).[9] We have performed calculations on $Ti(CO)_n$ complexes and confirmed the findings of ref. 7. We also performed calculations of Ti complexes and crystals with halogens and of Ti intercalants in solid Ca-doped titanium dioxide and Be-doped diamond where we also observed that the computed charge on Ti is relatively stable.

[a] Corresponding author. E-mail: mpemanzh@nus.edu.sg . Tel: +65 6516 4605; fax: +65 6779 1459.

The computed charges which are very different from formal oxidation states in magnitude and in change with the environment should already instill some doubt in the degree to which formal oxidation states reflect reality. To make sure this is not a fluke due to the method, we used in ref. 3 several *ab initio* methods which all told the same story. The work of ref. 7 used Mulliken and NBO charges. Perhaps there is a problem in how charges are defined in individual charge assignment schemes? In ref. 3, we directly analyzed valence electron density to ascertain that one electron charge resides on Ti in $TiO_2$ molecules and solids within less than half the Ti-O bond length, independently of definitions of common charge assignment schemes. We also confirmed that Bader charges provide a reasonable partitioning of space and Bader charges on the order of +2.5 |$e$| were most reasonable (among Bader, Mulliken, and Hirshfeld charges considered in ref. 3). We note that there is decent agreement with charge densities around Ti in a $TiO_6$ environment measured by X-ray/electron diffraction and computed by DFT.[10,11] We also note that Phillips fractional ionicity of $TiO_2$ (based on Pauling electronegativity[12]) is 0.6 which implies a significant covalent character (compared e.g. to 0.9 for LiF). Interestingly, latest ionic potential models[12] that are able to reproduce properties of various TiO$_2$ polymorphs use charges of +2.4 |$e$| and -1.2 |$e$| on Ti and O, respectively.[13,b]

If Ti$^{4+}$ generated by laser irradiation has a charge of +4 |$e$| and an oxidation state of +4, while Ti in TiO$_2$ retains one valence electron, how reasonable is it to persist using "Ti$^{4+}$" to describe TiO$_2$? One view justifying the use of Ti$^{4+}$ notation for Ti in TiO$_2$ is to acknowledge that the bonding has a substantial covalent character but justify Ti$^{4+}$ as a formal assignment. To argue that even if the density "leaks" onto Ti to "screen" the large ionic charge of +4 |$e$|, which would have been impossible to maintain unscreened (as some claim, although atomic charges of about +4 |$e$| have been reported in experimental and theoretical studies e.g. in sulfur compounds[14]), all valence electrons of Ti have been used to form bonds with O, and Ti is therefore fully oxidized, hence Ti$^{4+}$. There is reason in this argument; indeed, the one electron charge we found around Ti does not originate from a single-electron wavefunction (this can be surmised from the fact that it does not appear in spin density difference[15]). This argument, however, does not change the fact that there is a charge on Ti; nor is the oxidation state defined by the character of the underlying wavefunction. Indeed, the Merriam-Webster's Collegiate Dictionary defines oxidation number or state as "a positive or negative number that represents the effective charge of an atom or element that indicates the extent or possibility of its oxidation".[16] The "Ti$^{4+}$" language implies no further oxidation of Ti and no further reduction of O. But a degree of oxygen reduction is routinely seen in *ab initio* calculations of doped oxides including TiO$_2$[15] and oxygen redox has finally been embraced in such systems.[17] Are we going to write this off again as a fluke of computed charges such as Bader charges?

In semiconductors such as group IV monoelemental semiconductors as well as titania, vanadia etc. doped with interstitial alkali atoms, it is understood that the alkali atom donates its only valence electron to a state in the conduction band or in the gap[15,18,19] (or the valence band if the host is *p*-doped[20,21]) and is fully oxidized. The Bader charges do faithfully report (non-integer) values near +1 |$e$| on the alkali atom in these systems.[15,18-21] If the host is doped with an interstitial atom such as Mg, the Bader charge on Mg is not necessarily close to +2 |$e$| and may be close to +1.5 |$e$|.[20,21] However, the analysis of the density of states and spin polarization shows that two states become occupied in the conduction band of the host, or, if the host is *p*-doped, one state in the valence band and one in the conduction band.[20,21] That is how one can justify an assignment 1.5→2. Is there a way for a meaningful assignment 2.5→4 which, starting from Bader charges, would let one see donation of four electrons from Ti in a similar manner? This, we believe, would be a convincing argument in favor of $Ti^{4+}$. We have so far not been able to find such a way after performing calculations on Ti atoms in $TiCl_4$, $TiBr_4$ and $TiI_4$ in molecular and crystalline forms or interstitial Ti in Ca-doped $TiO_2$ and Be-doped diamond.

---

[b] Ref. 13 reported structures and elastic properties of rutile TiO$_2$ while we confirmed that a potential using these charges can reproduce structures of anatase, rutile and bronze TiO$_2$.

We thus have a situation where it is established that (i) TM oxides such as $TiO_2$ have a significant degree of covalent bonding; (ii) that static charges on atoms in them are not subject to quantization requirements which hold for long-range transferred charges; (iii) that physical charge around an atom is very different from and does not even significantly correlate with formal oxidation states which follow from a simple ionic approximation.

- Can one still assign integer charges via this ionic approximation or in a way that mimics its key properties? – Of course.
- Does one have to? – We do not think so.
- Should one? – We do not think so.
- Will such assignment reflect physical reality? - We do not think so. Or will it simply color the perception of it and create a comfortable fiction? - We think so. That it took so long to recognize oxygen reduction in metal ion batteries[17] may have to do with the ideological rigidity created by the ubiquitous use of the formal oxidation states to describe the mechanism of charge discharge.

We conclude based on the above that "$Ti^{3+}$" (if one wants to use an integer) is a better description of Ti in $TiO_2$ than "$Ti^{4+}$" which at this point looks rather unphysical. Paradigms tend to die slowly and painfully.[22] The one of formal oxidation states has been on life support for a number of years now[8,9] with no prospects of recovery; perhaps it is time to call the death panel and pull the plug.

**ACKNOWLEDGEMENTS**

This work was supported by the Ministry of Education of Singapore (Grant No. MOE2015-T2-1-011). We thanks Profs. Emily Carter, Paul Ayers, and Peter Deak for discussions and for referring us to some pertinent literature. We thank Prof. Aron Walsh for constructive criticism and some pointers that induced us to think deeper about Ti charge and oxidation state and the whole concept of formal oxidation states.